\documentclass[a4paper,11pt]{article}
\usepackage{amsfonts}
\usepackage{amssymb,epic,eepic} %
\author{Igor Bjelakovi\'c, Arleta Szko\l a\footnote{e-mail:\{igor, szkola\}@math.tu-berlin.de} \\
{\footnotesize Technische Universit\"at Berlin} \\
{ \footnotesize Fakult\"at II - Mathematik und Naturwissenschaften} \\
{ \footnotesize Institut f\"ur Mathematik MA 7-2} \\
{ \footnotesize Stra\ss e des 17. Juni 136} 
{ \footnotesize 10623 Berlin, Germany}}
\title{The Data Compression Theorem for Ergodic Quantum Information Sources}

\newcommand{\hr}{{\cal H}}

\newcommand{\nn}{{\mathbb N}}
\newcommand{\idn}{\mathbf{1}}
\newcommand{\zz}{{\mathbb Z}}

\newcommand{\eps}{{\varepsilon}}        

\newtheorem{theorem}{Theorem}[section]         
\newtheorem{proposition}[theorem]{Proposition} 

\begin{document}
\maketitle
\begin{abstract}
We extend the data compression theorem to the case of ergodic quantum
information sources.
 Moreover, we provide an asymptotically optimal
compression scheme which is based on the
concept of high probability subspaces. The rate of this compression
scheme is equal
to the von Neumann entropy rate.
\end{abstract}

\section{Introduction}
In classical information theory the Shannon entropy rate $h$ of discrete
stochastic processes modelling \emph{information sources} (IS) gives the
average information carried by individual signals: Operationally it
means that any  ergodic information source 
 can be compressed  by means of block
coding using asymptotically not more than
$h$ bits per signal in a way that there exist decompression alghorithms
with asymptotically  vanishing probability of  error. Using an
exponentially smaller
number of bits the compression/decompression algorithms will fail to be
asymptotically error-free.\\
In quantum information theory the corresponding quantity is the von
Neumann entropy rate $s$.
We show by construction (cf. Theorem \ref{compr_theo} below) that for any ergodic \emph{quantum information
source} (QIS) there exists an asymptotically reliable compression
scheme with rate $R$ equal to the von Neumann entropy rate
$s$. Here by rate we mean the asymptotic number of qubits
used per signal to represent the QIS. Designing compression algorithms
the goal is to achieve low rates.
It turns out that $s$ is the optimal rate in the
sense that a rate $R\geq s$ is a necessary condition on
asymptotical reliability of compression schemes and even more the fidelity of any compression
scheme with rate $R<s$  vanishes asymptotically. Of
course, this result depends on the underlying fidelity notion for the compression/decompression operations on the
QIS.
There are different definitions of
fidelity suited for different applications. In section \ref{sec_fidelity} we will discuss
some of them: the standard fidelity $F$ between two quantum states (it
can be seen as an extension of the overlap function of two pure states), the
ensemble fidelity $\bar F$ and the entanglement fidelity $F_{e}$. Our
result holds for
$\bar F$ as well as for $F_{e}$ .\\  
The main tool to construct compression
schemes achieving the optimal rate $s$ are high probability
subspaces. Compression maps that are essentially projections onto high
probability subspaces provide a solution to the problem of the optimal
data compression. A basic result
concerning high probability subspaces is proved in \cite{wir}. It asserts
the convergence of the
minimal logarithmic dimension rate of these subspaces to the von
Neumann entropy
rate $s$ in the case of ergodic QIS. This convergence was conjectured by Petz/Hiai
in \cite{petz}. \\ 
The concept of high probability subspaces
is crucial in the work of Petz/Mosonyi \cite{mosonyi}, where they prove a coding
theorem for the class of completely ergodic QIS.
Using projections onto high probability subspaces they show that completely ergodic QIS can be
compressed with any rate $R\geq s$ in such a way that the ensemble fidelity $\bar F$ is
asymptotically equal to $1$. On the other hand $\bar F$ cannot achieve
$1$ asymptotically if the rate satisfies $R<s$. 
The reason why they cannot conclude that in fact for $R<s$ the
asymptotical fidelity $\bar F$ is equal to $0$ is that they use the result of Hiai/Petz \cite{petz} which
provides bounds on limit superior and limit inferior and not the limit
of the minimal
logarithmic dimension rate of the high
probabilty subspaces. The result of  Petz/Mosonyi represents an
extension of the coding theorem formulated in \cite{jozsa} by Jozsa and
Schumacher  for the smaller class of
independent identically distributed (i.i.d.) QIS and proved in
\cite{jozsa} and \cite{barnum}. An analogous result for i.i.d. QIS using the entanglement
fidelity $F_{e}$ as a criterion for the reliability of
compression schemes is presented by Nielsen and Chuang in \cite{niel}.\\
In \cite{datta} Datta/Suhov treat the
case of certain weakly non-stationary quantum spin
systems. They show that, under the
condition of asymptotical reliability measured by the ensemble
fidelity $\bar F$, the optimal rate for compression
of information carried by Gibbs states of the considered interacting quantum spin
systems is given by the von Neumann entropy rate.

\section{Quantum Information Sources}

Classical discrete IS are stochastic processes, i.e. sequences of
random variables $\{X_{i}\}_{i\in \zz}$ with a joint distribution $P$,
each random variable $X_{i}$ taking values from a set $A$ called
alphabet. We will consider only the case of IS over finite alphabets. A possible
realisation of an (discrete) IS over an alphabet $A$ is a physical process
producing at discrete times physical systems with identical state spaces equal to
$A$ and the individual states of the systems being random variables
taking values from $A$  according to a probability rule $P$ determined
by the given IS. Alternatively a classical IS can be viewed as a classical spin
chain possibly coupled to an external enviroment. An IS is the first
stage in the process of information transmission or storage, it
provides information which is sent via a classical channel.  \\
Equivalent to the
stochastic process model is the Kolmogorov representation of IS
(cf. \cite{shields}). It
describes an IS as a dynamical system $(
A^{\infty}, \frak{A}^{\infty}, \nu, T)$ on a
doubly infinite product space $A^{\infty}:= \dots \times A \times A
\times \dots$, where $\frak{A}^{\infty}$ is the $\sigma$-field
generated by cylinder sets,  $T$ is the shift on $A^{\infty}$ and
$\nu$ a probability measure on  $(A^{\infty}, \frak{A}^{\infty})$ uniquely determined by the
probability distribution $P$ of the stochastic process.   \\
A discrete QIS can be viewed as a quantum spin chain
possibly coupled to an external environment where one focuses on the
information carried by the quantum state. Alternatively, we can think of
a QIS as a device that sends quantum physical systems of a
fixed type, prepared in a joint generally mixed and entangled
state. In both cases, 
a QIS provides input for quantum channels. \\
Before we present a mathematical model for QIS, which
corresponds to the Kolmogorov representation of classical IS, we
introduce quasilocal $C^{*}$-algebras as the non-commutative
counterpart of the doubly infinite product space $A^{\infty}$. 
This standard mathematical formalism is
introduced in detail e.g. in  \cite{ruelle}. One starts with the group
$\zz$. To each $z \in \zz$ there is associated  a $C^{*}$-algebra
${\cal A}_{z}$.  Each ${\cal
  A}_{z}$ is isomorphic
to a fixed finite dimensional unital $C^{*}$-algebra 
${\cal A}$, which is in general non-commutative. The $C^{*}$-algebra ${\cal A}$ corresponds to an algebra
of observables of a quantum system and the isomorphism between the
${\cal A}_{z}$ reflects the assumption that the source is
emitting quantum systems of a fixed type. \\
In this paper we will be mainly concerned with the case ${\cal
  A}={\cal B}(\hr)$, the linear operators on the finite dimensional
Hilbert space $\hr$.
For a finite subset $\Lambda \subset
\zz $ the algebra ${\cal
  A}_{\Lambda}$   is given by ${\cal A}_{\Lambda}:= \bigotimes_{z \in \Lambda} {\cal
  A}_{z}$. The quasilocal $C^{*}$-algebra ${\cal
  A}^{\infty}$ is defined as the operator norm closure of the local
$^{*}$-algebra ${\cal A}_{\mathrm{loc}}:=\bigcup_{\Lambda \subset \zz}{\cal
  A}_{\Lambda}$.\\
A state on the quasilocal algebra is given by a normed positive functional $\Psi$, i.e. $\Psi(\idn)=1$ and $\Psi(A)\geq 0$ for all $
A\in {\cal A}^{\infty} $ with $A\geq 0$ . There is one-to-one
correspondence between the state $\Psi$
and a consistent family of states
$\{\Psi^{(\Lambda)}\}_{\Lambda \subset \zz}$, where each
$\Psi^{(\Lambda)}$ is the restriction of $\Psi$ to the finite dimensional subalgebra ${\cal
  A}_{\Lambda}$ of ${\cal A}^{\infty}$ and consistency means that $\Psi^{(\Lambda)}=\Psi^{(\Lambda^{'})}\upharpoonright{\cal
  A}_{\Lambda}$ for $\Lambda \subset
\Lambda^{'}$. For most purposes it suffices to deal only with
integer intervals $\Lambda=\{z_{1},z_{1}+1,\ldots ,z_{2}\}$ with
integers $z_{1}\leq z_{2}$. For each $\Psi^{(\Lambda)}$ there exists a unique density operator
$\rho^{(\Lambda)}\in {\cal A}_{\Lambda}$,
such that $\Psi^{(\Lambda)}(a)=\textrm{tr}_{\Lambda}\rho^{(\Lambda)}a,\ a
\in {\cal A}_{\Lambda}$ and $\textrm{tr}_{\Lambda}$ is the trace on
${\cal A}_{\Lambda}$. \\
On ${\cal A_{\mathrm{loc}}}$ we define the shift $T$ that
acts in in the following way:
For integers $z_{1}\leq z_{2}$ and $\Lambda=\{z_{1},z_{1}+1,\ldots
,z_{2}\} $ we have 
\begin{eqnarray*}
T:{\cal A}_{\Lambda}\to {\cal A}_{\Lambda +1},\ a\simeq
a\otimes\idn\mapsto T(a)=\idn\otimes a\simeq a.
\end{eqnarray*}
The canonical extension of $T$ onto ${\cal A^{\infty}}$ is an 
$*$-automorphism on ${\cal A^{\infty}}$ and the integer powers of $T$,
$\{T^{z}\}_{z \in \zz}$,
represent an action of the translation group $\zz$ by
automorphisms on ${\cal A}^{\infty}$. The triple $({\cal A}^{\infty},
\Psi, T)$ defines a quantum dynamical system.

The mathematical model for a discrete QIS
is a quantum dynamical system
$({\cal A}^{\infty}, \Psi, T)$, where ${\cal A}^{\infty}$ is a
quasilocal $C^{*}$-algebra over $\zz$ constructed from a
finite dimensional $C^{*}$-algebra ${\cal A}$, $\Psi$ is a
state and $T$ the shift on ${\cal A^{\infty}}$.\\ \\
\textbf{Remark:}
\emph{If ${\cal A}$ is an \emph{abelian} finite dimensional
  $C^{*}$-algebra then by the Gelfand isomorphism ${\cal A}$ can be
  identified with $C(A)$, the algebra of functions on a set
  $A$ with $|A|= \dim{\cal A}$ and ${\cal A}^{\infty}$ is
  $^{*}$-isomorphic to $C(A^{\infty})$. Further by the Riesz representation
  theorem there exists a probability measure $\nu$ on
  $(A^{\infty}, \frak{A}^{\infty})$ uniquely determined by $\Psi(a)=\sum_{\mathbf{i}\in
    A^{\Lambda}}a(\mathbf{i})\nu^{(\Lambda)}(\mathbf{i})$, for all $a\in
  {\cal A}_{\Lambda}$ and arbitrary $\Lambda \subset \zz$, where
  $a(\cdot) \in C(A^{\Lambda})$ is the Gelfand representation of $a \in
  {\cal A}_{\Lambda}$. Then the IS given by $({\cal A}^{\infty},
  \Psi, T)$ has the Kolmogorov representation  $(A^{\infty},
  \frak{A}^{\infty},  \nu, T)$. Hence, this construction leads back to
the classical IS.}  \\ \\
For simplicity in the following sections we will restrict to the case ${\cal A}=
{\cal B}(\hr)$.\\ \\
$({\cal A}^{\infty}, \Psi, T)$ is a
\emph{stationary} QIS if for all $a \in {\cal A}^{\infty}$:
\begin{eqnarray}
\Psi(Ta)=\Psi(a).
\end{eqnarray}
As we will deal only with stationary QIS, we will assume without
loss of generality that all integer intervals are of the form
$\Lambda=\{1,\ldots ,n\}$ with $n\geq 1$. Furthermore we write $\rho^{(n)}$
instead of $\rho^{(\Lambda)}$ and $\Psi^{(n)}$ instead of $\Psi^{(\Lambda)}$.\\   
A stationary QIS $({\cal A}^{\infty}, \Psi, T)$ is \emph{ergodic} if
\begin{eqnarray*}
\lim_{n \to \infty}\Psi\left((\frac{1}{n}\sum_{i=0}^{n-1}T^{i}(a))^{2}\right) = \Psi(a)^{2}
\end{eqnarray*} 
holds for all self-adjoint $a \in {\cal A^{\infty}}$. It is important
to realize that for a QIS as a dynamical system on a quasilocal algebra this definition of ergodicity is equivalent to the
definition used in \cite{wir} (cf. Proposition 6.3.5 in \cite{ruelle}), where $({\cal A}^{\infty}, \Psi, T)$ is
said to be
ergodic if $\Psi$ is an extremal point in the compact convex set of
stationary states on ${\cal A}^{\infty}$, (cf. \cite{israel}). Thus
the results for ergodic quantum dynamical systems presented in
\cite{wir} hold for the ergodic QIS defined
above.\\
Finally in this section we introduce the entropy rate $s(\Psi)$ of a
stationary QIS $({\cal A}^{\infty}, \Psi, T)$, which
is the crucial quantity in the present paper. Recall the one-to-one
correspondence of a stationary $\Psi$ on ${\cal A}^{\infty}$ and the
family of density operators $\{\rho^{(n)}\}n \in \nn$. The \emph{entropy
rate} $s(\Psi)$ is then defined by
\begin{eqnarray}\label{entropyrate}
s(\Psi):=\lim_{n\to \infty}\frac{1}{n}S(\rho^{(n)}),
\end{eqnarray}
where $S(\rho^{(n)}):= -\textrm{tr}\rho^{(n)}\log_{2} \rho^{(n)}$ is the
von Neumann entropy of the density operator $\rho^{(n)}$. 


\section{Data Compression Schemes}
In order to define lossless data compression schemes for
encoding quantum signals we need the concept of  \emph{trace preserving quantum operations}.
A physical approach to trace preserving quantum operations can
be obtained as follows. Consider a quantum system $S$ prepared in some
state $\rho$ acting on the Hilbert space $\hr$. We imagine that this
system interacts with its enviroment, a quantum system $S_{\mathrm{env}}$ in a
state $\rho_{\mathrm{env}}$ on the finite dimensional Hilbert space $\hr_{\mathrm{env}}$.
 The system $S \times S_{\mathrm{env}}$ is closed and we make the assumption that it is
initially in the product state $\rho\otimes \rho_{\mathrm{env}}$ on
$\hr\otimes\hr_{\mathrm{env}}$. As a state of a closed system it
undergoes a unitary evolution represented by a unitary operator
$U$ on $\hr\otimes\hr_{\mathrm{env}}$. The corresponding evolution of the state
$\rho$ of $S$ is usually not unitary, i.e. irreversible. It is given by a trace preserving
quantum operation ${\cal E}$:
\begin{eqnarray}\label{operation}
 {\cal E}(\rho):=\textrm{tr}_{\hr_{\mathrm{env}}}(U(\rho\otimes\rho_{\mathrm{env}})U^{\ast}). 
\end{eqnarray}
 It can be shown that each trace preserving quantum operation ${\cal E}$ posseses the following
representation known as Kraus or sum representation
(cf. \cite{hellwig-kraus1}, \cite{hellwig-kraus2}, \cite{niel})
\begin{eqnarray*}
  {\cal E}(\rho)=\sum_{i}E_{i}\rho E_{i}^{\ast},
\end{eqnarray*}
where $E_{i}\in {\cal B}(\hr)$ and
$\sum_{i}E_{i}^{\ast}E_{i}=\idn$. This description contains, for
example, the cases of the unitary time evolution and general measurements.
We remark that trace preserving quantum operations may be described in a more
elegant way within the framework of completely positive linear
maps between $C^{\ast}$-algebras (cf. \cite{davies}, \cite{kraus}).\\
\\
A \emph{compression scheme} $({\cal C},{\cal D})$ for stationary QIS
is a sequence $\{ ({\cal C}^{(n)}, {\cal D}^{(n)})\}_{n \in \nn} $
 of pairs of trace preserving quantum operations 
\begin{eqnarray}\label{compr_scheme}
{\cal C}^{(n)}: {\cal S}(\hr ^{\otimes n}) \longrightarrow {\cal
  S}(\hr^{(n)}), 
\end{eqnarray}
\begin{eqnarray*}
{\cal D}^{(n)}: {\cal S}(\hr ^{(n)}) \longrightarrow {\cal
  S}(\hr^{\otimes n})  
\end{eqnarray*}
where $\hr^{(n)} \subseteq \hr^{\otimes n}$ for
all $n \in \nn$ and ${\cal S}(\hr ^{\otimes n}),\ {\cal S}(\hr ^{(n)})
$ denote the sets of density operators on $\hr^{\otimes n}$ resp. $
\hr^{(n)}$. We refer to ${\cal C}^{(n)}$, ${\cal D}^{(n)}$ as compression
resp. decompression map.\\
The \emph{rate} $R({\cal C})$ of a compression scheme
$({\cal C},{\cal D})$
is defined by
\begin{eqnarray}\label{comp_rate}
R({\cal C}):= \limsup _{n \to \infty} \frac{\log_{2} \dim \hr^{(n)}}{n}.
\end{eqnarray}


\section{Fidelities}\label{sec_fidelity}

In this section we review the basic notions and properties of the fidelity and derived
quantities needed to measure the distance between two quantum states. 
The fidelity $F$ between two density operators $\rho$ and $\sigma$ acting on some
finite dimensional Hilbert space $\hr$ is defined by
\begin{equation}\label{eq:fidelity}
 F(\rho,\sigma):=\textrm{tr}\sqrt{\sqrt{\rho}\sigma \sqrt{\rho}}.
\end{equation}
The fidelity is symmetric in its
entries and takes values between $0$ and $1$ with $ F(\rho,\sigma)=0$
iff $\rho$ and $\sigma$ are supported on orthogonal subspaces.
$ F(\rho,\sigma)=1$ appears only in the case $\rho =\sigma$.
In light of these properties it is reasonable to interpret
the fidelity as a measure of distinguishability of two 
density operators which reduces to the well known overlap 
 $|\langle \psi | \phi\rangle|$ in the case of
pure states $| \psi \rangle \langle \psi |$ and 
$| \phi \rangle \langle \phi |$ on $\hr$.
 Moreover $F$ is jointly concave and increasing under trace 
preserving quantum 
operations. The proofs of these facts may be found in \cite{niel}.
The fidelity $F$ is equivalent to the familiar
trace distance of two density operators in the following sense:
\begin{equation}\label{eq:estimate}
1-F(\rho,\sigma)\leq\frac{1}{2}\textrm{tr}|\rho -\sigma|\leq
\sqrt{1-(F(\rho,\sigma))^{2}} \quad(\textrm{cf. ~\cite{niel}}).  
\end{equation}
But the trace distance can be represented as (cf. \cite{niel})
\begin{displaymath} 
\frac{1}{2}\textrm{tr}|\rho -\sigma|=\max\{\textrm{tr}(P(\rho -\sigma)):P=P^{\ast}=P^{2}\}.
\end{displaymath}
This equality has the following meaning: the orthogonal
projections appearing in the above equation are usualy interpreted 
as ideal ``yes-no'' measurements. The outcome ``yes'' (resp. ``no'')
is  represented by $P$ (resp. $ \idn -P$). The trace distance quantifies 
the largest difference of probabilities for obtaining outcome
``yes'' if we perform measurements on quantum systems in the states
$\rho$ and $\sigma$. This relation between the fidelity and the trace
distance gives us an idea about the operational
interpretation of the fidelity. 

The question how well is the state of the open quantum system preserved by
a time evolution, a measurement or more generally by an arbitrary quantum
processes defined by a Kraus representation leads to several fidelity
concepts. The first one is the \emph{entanglement fidelity}
$F_{e}$ which is a function 
of a density operator $\rho$ and a quantum operation ${\cal
  E}$. It is defined by
\begin{eqnarray}\label{F_e_def}
F_{e}(\rho, {\cal E}):= \left( F(|\Psi \rangle \langle \Psi |,(\mathbf{1}
\otimes {\cal E})(|\Psi \rangle \langle \Psi |))\right)^{2},
\end{eqnarray} 
where $|\Psi \rangle \in \hr^{'} \otimes \hr$ is an arbitrary
purification of $\rho$, i.e. $\textrm{tr}_{\hr^{'}}|\Psi \rangle
\langle \Psi |= \rho$. It can be shown that this definition does not
depend on the particular choice of the purification of $\rho$, cf. \cite{niel}.
\\ Let  ${\cal E}(\rho)=\sum_{i}E_{i}\rho E_{i}^{*}$   
be the sum
representation of ${\cal E}$ for all
density operators $\rho$ on $\hr$, i.e. $E_{i}\in {\cal
  B}(\hr)$ and $\sum_{i}E_{i}^{*} E_{i}=
\textbf{1}$. Then it holds
\begin{eqnarray}\label{F_e_sum}
F_{e}(\rho, {\cal E})=\sum_{i}|\textrm{tr}\rho E_{i}|^{2}.
\end{eqnarray}   
This formula implies  that the entanglement fidelity is a
convex function of the density operator. Indeed, the last expression 
is merely the squared norm of a complex vector with the components
$\textrm{tr}(\rho E_{i})$, which depend affinely on $\rho$. 
Moreover, every norm is a convex function, so we obtain the claimed
convexity of $F_{e}$. The intuition behind the definition
(\ref{F_e_def}) is that what we want to preserve is the purifications of 
a given state. If the state is mixed then all purifications are
entangled pure states.

In order to define the \emph{ensemble fidelity} $\bar{F}$ we start with a
finite set of symbols $\{1,\ldots ,n\}$ (a classical alphabet)
which are drawn according to a probability distribution
$(p_{1},\ldots ,p_{n})$. We associate to this set of symbols
a fixed set of density operators $\{\rho_{1},\ldots \rho_{n}\}$ on
$\hr$ and define the ensemble fidelity by
\begin{equation}\label{eq:ensemble}
 \bar{F}(\{(p_{i},\rho_{i})\}_{i=1}^{n},{\cal E}):=\sum_{i=1}^{n}p_{i}(F(\rho_{i},{\cal E} (\rho_{i})))^{2}, 
\end{equation}
where ${\cal E}$ is a quantum operation. The weighted ensemble of $n$
quantum states  $\{(p_{i},\rho_{i})\}_{i=1}^{n}$ represents a convex
 decomposition of the density operator
 $\rho=\sum_{i=1}^{n}p_{i}\rho_{i}$. If the $\rho_{i}$ are all pure states
 then we call the ensemble or the convex decomposition a pure one. We will
 denote by $F_{s}( \rho, {\cal E})$ the supremum over pure convex
 decompositions
of the ensemble fidelities for a  density operator $\rho$ and a
quantum operation ${\cal E}$: 
\begin{eqnarray}\label{eq:supfid}
F_{s}( \rho, {\cal E}):= \sup
\{ \bar{F}(\{(p_{i},P_{i})\}_{i=1}^{n},{\cal E})&:&
\{(p_{i},P_{i})\}_{i=1}^{n} \textrm{ pure convex} \nonumber \\ & &  \textrm{decomposition of } \rho\}.
\end{eqnarray}
The idea behind the definition (\ref{eq:ensemble}) is that the classical
alphabet is represented by quantum systems prepared in 
the states from some fixed set. For example we can encode 
the alphabet $\{0,1\}$ into two different polarization directions
of photons. The probability of occurence of each polarisation direction
is determined by the probability distribution on the classical
alphabet. The ensemble fidelity  $\bar{F}$ appears mainly in problems
concerning classical information to be e.g. stored on or transmitted via quantum states. 

We conclude this section with a basic relation among 
several notions of fidelity introduced here. 
For a fixed density operator $\rho$ we define
\begin{eqnarray*}
\bar \textrm{F}_{\rho, {\cal E}}:= \{\bar F(\{(p_{i}, \rho_{i})\}_{i}, {\cal
  E})| \sum_{i}p_{i}\rho_{i}=\rho\}.
\end{eqnarray*}
It holds
\begin{equation}\label{eq:relation}
0\leq F_{e}(\rho ,{\cal E})\leq  \bar F \leq F(\rho ,{\cal
  E}(\rho))\leq 1,\quad \bar F \in \bar \textrm{F}_{\rho, {\cal E}}.
\end{equation}
The second inequality follows immediately from the convexity
of the entanglement fidelity. The third inequality
holds because the fidelity $F$ is jointly concave. Observe that according
to (\ref{eq:relation}) we can give upper and lower
bounds for $\bar F$ which depend exclusively on the density operator
$\rho$ corresponding to the convex decomposition in consideration.
The inequality (\ref{eq:relation}) will play a crucial role in our
derivation of data compression theorem.


\section{Data Compression Theorem}

One of the interests in the quantum information theory is an economical
and errorfree
storage or transmission of quantum information. In other words the question is: what is the minimal amount of
resources measured in units of qubits or equivalently in Hilbert space dimensions needed to store
quantum states faithfully? This question has been resolved in the case
of memoryless sources using the entanglement fidelity $F_{e}$ as a criterion
for reliability,  \cite{niel}:
Each compression scheme possesing a rate smaller than the
von Neumann entropy rate cannot be reliable in the sense that the
entanglement fidelity tends to $0$. 
 It has been shown in \cite{barnum} that for encodings of classical
memoryless sources into some fixed set of pure quantum states, as
described in the previous section, an analogous assertion holds. In this
case the reliability is measured by the ensemble fidelity $\bar F$.
In both cases compression schemes have been constructed with  rates, that can be made arbitrary close to the von Neumann entropy $S(\rho)$. An essential ingredient was  the quantum asymptotic
 equipartition property (AEP) for memoryless QIS.
An extension of the quantum AEP to the more general case
of ergodic QIS was formulated and proved in \cite{wir}.
\begin{theorem}[Quantum AEP Theorem]\label{SM-theo}
Let $({\cal A}^{\infty}, \Psi, T)$ be an ergodic quantum information
source with the entropy rate $s(\Psi)$ defined by eqn. (\ref{entropyrate}). Then for any $\eps>0$ there
exists an $N_{\eps}\in \nn$ such that for all $n\geq N_{\eps}$ there
exists a subspace
${\cal T}^{(n)}_{\eps}\subseteq \hr^{\otimes n}$ such that

\begin{itemize}
\item[1)] $\textup{tr}(\rho^{(n)} P_{{\cal
    T}^{(n)}_{\eps}})\geq 1-\eps$, where $P_{{\cal
    T}^{(n)}_{\eps}}$ is the projector onto the subspace
${\cal T}^{(n)}_{\eps}$,

\item[2)] 
$2^{n(s(\Psi)-\eps)}\leq \textup{tr}(P_{{\cal
    T}^{(n)}_{\eps}})\leq 2^{n(s(\Psi)+\eps)}$.

\end{itemize}
Moreover, these subspaces can be chosen as  
\begin{eqnarray}\label{typ_subspace} 
{\cal T}^{(n)}_{\eps}:=&\textup{span}\{e_{i}^{(n)}\in \hr^{\otimes n} |&
\textup{tr}(\rho^{(n)}P_{ e_{i}^{(n)}})\in [2^{-n(s(\Psi)+
  \eps)},2^{-n(s(\Psi)- \eps)} ], \nonumber \\ & & e_{i}^{(n)} \textrm{eigenvector of } \rho^{(n)}
\}. 
\end{eqnarray}
\end{theorem}
\textbf{Remark:}
\emph{The above theorem represents a simplified form of the Quantum Shannon-McMillan
Theorem presented in \cite{wir} for the more general case of higher
dimensional quantum dynamical lattice systems. Furthermore it is important to
notice that the version  presented in \cite{wir} also includes the case of
${\cal A}^{\infty}$ constructed from a subalgebra ${\cal A}\subseteq
{\cal B}(\hr)$. If the subalgebra ${\cal A}$ is commutative then the
Quantum Shannon-McMillan Theorem coincides with the classical
theorem.}\\ \\  Each of the ${\cal T}_{\eps}^{(n)} \subseteq \hr^{\otimes n}$ defined by
(\ref{typ_subspace}) represents a subspace of probability close to $1$ for
large $n$ in the case of an ergodic source $({\cal A}^{\infty}, \Psi,
T )$. In analogy to the classical theory we
will call such a space the  $\eps$-typical subspace of
$\hr^{\otimes n}$ with respect to $\rho^{(n)}$.\\
\\ The next proposition is
strongly related to the Quantum AEP Theorem. In \cite{wir} it is
proven for the higher dimensional case. The proposition is 
crucial for the proof of the second and third part of the Data Compression
Theorem presented below. These parts say that an asymptotically
reliable compression to the von Neumann entropy rate is achievable and is an
optimal one.
The relevant quantity for compression is the minimal logarithmic dimension of
subspaces of $\hr^{\otimes n}$ depending on the minimal required probability of the subspaces.
\begin{eqnarray*}
\beta_{\eps, n}(\Psi):= \min \{ \log_{2} ( \textrm{tr} q) | \ q \in {\cal
  B}(\hr^{\otimes n}) \ \textrm{projector}, \ \textrm{tr}\rho^{(n)}q
\geq 1-\eps\},\ \eps \in (0, 1).
\end{eqnarray*}
We refer to  subspaces ${\cal P}^{(n)}_{\eps} \subset \hr^{\otimes n}$ with
$\textrm{tr}\rho^{(n)}P_{{\cal P}^{(n)}_{\eps}}\geq 1-\eps$ and 
$\log (\textrm{tr}P_{{\cal P}^{(n)}_{\eps}})=\beta_{\eps, n}(\Psi) $ as
  high probability subspaces (with resp. to $\rho^{(n)}$)
  corresponding to the level $\eps$, cf. \cite{mosonyi}.\\
It turns out that the asymptotic rate of $\beta_{\eps, n}(\Psi)$ does
not depend on $\eps$ in the case of an ergodic state $\Psi$ and is equal to the entropy
rate $s(\Psi)$. 
\begin{proposition}\label{beta=s}
Let \(( {\cal A}^{\infty}, \Psi, T)\) be an ergodic quantum
information source with the
entropy rate \(s(\Psi)\). Then for every $\eps
\in (0,1)$
\begin{eqnarray}\label{lim_beta=s}
\lim_{n \to \infty}\frac{1}{n} \beta_{\eps,n}(\Psi)=s(\Psi).
\end{eqnarray}
\end{proposition}
Now, disposing
 of the above proposition we can extend results concerning compressibility of information to the case of
 correlated (ergodic) QIS. 
\begin{theorem}[Data Compression Theorem]\label{compr_theo}
Let $( {\cal A}^{\infty}, \Psi, T)$ be an ergodic quantum information source
with the entropy rate $s(\Psi)$.
\begin{itemize}
\item[1)]  Each compression scheme $({\cal C},{\cal
    D})$ satisfying 
\begin{eqnarray}\label{limit}
\lim_{n\to \infty} \bar F ( \{ ( \lambda_{i}^{(n)}, P_{i}^{(n)} )
\}_{i=1}^{k_{n}}, {\cal D}^{(n)}\circ {\cal
  C}^{(n)} )=1
\end{eqnarray}
for some sequence
$\{\{(\lambda_{i}^{(n)},P_{i}^{(n)})\}_{i=1}^{k_{n}}\}_{n\in\nn}$ of pure
convex decompositions of $\rho^{(n)}$, respectively,
fulfils
\begin{eqnarray*}\label{compr_theo_0} 
R({\cal C})\geq s(\Psi).
\end{eqnarray*}
\item[2)] There
exists a compression scheme $ ({\cal C},{\cal D})$ with 
$R({\cal C})=s(\Psi)$ such that
\begin{eqnarray*}\label{compr_theo_1}
\lim _{n \to \infty} F_{e}(\rho^{(n)}, {\cal D}^{(n)}\circ {\cal
  C}^{(n)})=1.
\end{eqnarray*}
\item[3)] Any compression scheme $({\cal C},{\cal D})$ with $R({\cal C})<s(\Psi)$ satisfies

\begin{eqnarray}\label{compr_theo_2}
\lim_{n \to \infty}F_{s}(\rho^{(n)}, {\cal D}^{(n)}\circ {\cal C}^{(n)})=0,
\end{eqnarray}
where $F_{s}$ is defined by (\ref{eq:supfid}).

\end{itemize}
\end{theorem}
Taking into account the relation (\ref{eq:ensemble})
we use different notions of fidelity in the seperate parts of
the above theorem. In this way we obtain that the von Neumann entropy
rate is the optimal compression rate using the fidelity $\bar F$ as well as 
$F_{e}$.\\ 
It should be helpful to sketch the ideas which lead to the proof of
the theorem above. The first item in the theorem is essentially a
consequence of the monotonicity of the relative entropy
(cf. \cite{uhlmann}) and the Fannes inequality (cf. \cite{fannes})
modulo some elementary estimates. The second item is derived from the fact stated in the
Proposition \ref{beta=s} saying that the asymptotic rate of $\beta_{\eps,n}$
is given by the von Neumann entropy rate and does not depend on the
level $\eps$.   
Compression schemes $({\cal C},{\cal D})$ consisting of compression maps which are
essentially projections
onto the high probability subspaces and the canonical embeddings as
decompression maps posses a rate equal to the von Neumann entropy
rate. 
So, if we combine appropriately high probability subspaces such that
their corresponding
levels tend to $0$, we can achieve that the
entanglement fidelity becomes arbitrary close to $1$. This strategy
leads directly to the proof of the second part of the
theorem. Finally, the third item in the above theorem can be proved
using the fact that $F_{s}$ is bounded from above by the maximal
expectation value of projectors $P \in \hr^{\otimes n}$ satisfying the dimension
condition $\textrm{tr}P= \dim \hr ^{(n)}$. But if the rate of a data compression scheme is asymptotically
smaller than the von Neumann entropy rate then according to the
Proposition \ref{beta=s} the  expectation values of projectors
providing the upper bounds for $F_{s}$
must vanish asymptotically.  \\
\textbf{Proof of Theorem \ref{compr_theo}:}\emph{ Proof of 1)}  Fix a
convex decomposition of $\rho^{(n)}$ into one dimensional  projectors
$\{P_{i}^{(n)}\}_{i=1}^{k_{n}}$ corresponding to the set of weights 
$\{{\lambda}_{i}^{(n)}\}_{i=1}^{k_{n}}$. Following an idea of
M. Horodecki in \cite{horodecki} we arrive at the following
elementary inequalities using the relative entropy and its decreasing 
behaviour with respect to the trace preserving operations (cf. \cite{uhlmann}):
\begin{eqnarray}
\log_{2} \dim \hr^{(n)}& \geq & S({\cal
  C}^{(n)}(\rho^{(n)}))\nonumber \\
&\geq & S({\cal
  C}^{(n)}(\rho^{(n)}))-\sum_{i=1}^{k_{n}}{\lambda}_{i}^{(n)}S({\cal
  C}^{(n)}
(P_{i}^{(n)}))\nonumber \\
&= & \sum_{i=1}^{k_{n}}{\lambda}_{i}^{(n)}S({\cal
  C}^{(n)}(P_{i}^{(n)}),{\cal C}^{(n)}(\rho^{(n)}))\nonumber \\ 
& \geq & \sum_{i=1}^{k_{n}}{\lambda}_{i}^{(n)}S({\cal D}^{(n)}\circ{\cal
  C}^{(n)}(P_{i}^{(n)}),{\cal D}^{(n)}\circ{\cal C}^{(n)}(\rho^{(n)}))\nonumber \\
&= & S({\cal D}^{(n)}\circ{\cal
  C}^{(n)}(\rho^{(n)}))-\sum_{i=1}^{k_{n}}{\lambda}_{i}^{(n)}S({\cal D}^{(n)}\circ{\cal
  C}^{(n)}
(P_{i}^{(n)}))\nonumber
\end{eqnarray}
In the next step we will show that
\begin{eqnarray}\label{entropy}
  \lim_{n\to \infty}\frac{1}{n}S({\cal D}^{(n)}\circ{\cal
  C}^{(n)}(\rho^{(n)}))=s(\Psi),
\end{eqnarray}
and
\begin{eqnarray}\label{zero}
\lim_{n\to\infty}\frac{1}{n}\sum_{i=1}^{k_{n}}{\lambda}_{i}^{(n)}S({\cal D}^{(n)}\circ{\cal
  C}^{(n)}
(P_{i}^{(n)}))=0,
\end{eqnarray}
holds, which implies the first part of the theorem. 
By (\ref{eq:estimate}) and the Fannes inequality (cf. \cite{fannes}) we have
\[\frac{1}{n}|S(\rho^{(n)})-S({\cal D}^{(n)}\circ{\cal C}^{(n)}(\rho^{(n)}))|\leq
2\log_{2} d\sqrt{1-(F( \rho^{(n)},{\cal D}^{(n)}\circ{\cal C}^{(n)}(\rho^{(n)})))^{2}}+\frac{1}{n}.\]
Employing the limit assertion (\ref{limit}) and joint concavity of the fidelity 
we obtain (\ref{entropy}).\\
Fix $\eps\in (0,1)$.  We consider the set
\[ A_{\eps}^{(n)}:=\{i\in\{1,\ldots,k_{n}\}|\ (F(P_{i}^{(n)},{\cal D}^{(n)}\circ{\cal
  C}^{(n)}
(P_{i}^{(n)})))^{2}<1-\eps\}\] and estimate 
\begin{eqnarray}\label{abschatzung}
\sum_{i=1}^{k_{n}}\lambda_{i}^{(n)}F^{2}(P_{i}^{(n)},{\cal D}^{(n)}\circ{\cal
  C}^{(n)}
(P_{i}^{(n)}))\leq (1-\eps)\sum_{i\in A_{\eps}^{(n)}}\lambda_{i}^{(n)}+
\sum_{i\in A_{\eps}^{(n)c}}\lambda_{i}^{(n)},
\end{eqnarray}
where $A_{\eps}^{(n)c}$ denotes the complement of $A_{\eps}^{(n)}$. We
claim that for all $\eps\in (0,1)$
\begin{eqnarray}\label{lowfid}
\lim_{n\to\infty}\sum_{i\in A_{\eps}^{(n)}}\lambda_{i}^{(n)}=0.   
\end{eqnarray}
In fact, suppose that for some $\eps\in (0,1)$
\[\limsup_{n\to\infty}\sum_{i\in A_{\eps}^{(n)}}\lambda_{i}^{(n)}=a>0.\]
Then there would exist a subsequence, which we denote again by
$\{A_{\eps}^{(n)}\}_{n\in\nn}$ for simplicity, with
\[\lim_{n\to\infty}\sum_{i\in A_{\eps}^{(n)}}\lambda_{i}^{(n)}=a.\]
After taking
limits in (\ref{abschatzung}) this would imply the following contradictory inequality 
\[ 1\leq (1-\eps)a + (1-a).\]
By (\ref{lowfid}), it suffices to show that
\[ \lim_{n\to\infty}\frac{1}{n}\sum_{i\in A_{\eps}^{(n)c}}{\lambda}_{i}^{(n)}S({\cal D}^{(n)}\circ{\cal
  C}^{(n)}
(P_{i}^{(n)}))=0.\]
For small $\eps\in (0,1)$ and for $n$ large enough we have
\begin{eqnarray*}
\frac{1}{n}\sum_{i\in A_{\eps}^{(n)c}}{\lambda}_{i}^{(n)}S({\cal
  D}^{(n)}\circ {\cal C}^{(n)}
(P_{i}^{(n)}))&\leq& \frac{1}{n}\sum_{i\in A_{\eps}^{(n)c}}{\lambda}_{i}^{(n)}(2n\log_{2}(d) \sqrt{\eps}+1)\\
&\leq& 2\log_{2}(d) \sqrt{\eps}+\frac{1}{n},
\end{eqnarray*}
where in the first inequality we have applied Fannes inequality to the expressions
$S({\cal
  D}^{(n)}\circ {\cal C}^{(n)}
(P_{i}^{(n)}))=|S(P_{i}^{(n)})- S({\cal
  D}^{(n)}\circ {\cal C}^{(n)}
(P_{i}^{(n)}))|$, respectively.  Since
$\eps$ can be made arbitrarily small, we have
\[\lim_{n\to\infty}\frac{1}{n}\sum_{i\in A_{\eps}^{(n)c}}{\lambda}_{i}^{(n)}S({\cal
  D}^{(n)}\circ {\cal C}^{(n)}
(P_{i}^{(n)}))=0.\]
\emph{Proof of 2)} By proposition \ref{beta=s} we have
\[R({\cal C})= \lim_{n \to \infty} \frac{\log_{2} \dim {\cal
    P}_{\eps}^{(n)}}{n} = \lim_{n \to \infty} \frac{1}{n} \beta_{\eps,n}=s(\Psi) \]
for each $\eps \in (0,1)$. A simple 
argument shows that there exists a sequence $\eps_{n}\searrow 0$, for $
n\to \infty $, such that
\[\lim_{n\to \infty}\frac{1}{n}\beta_{\eps_{n},n}=s(\Psi).\]
 We consider the compression scheme $({\cal C}, {\cal D})$, where
for each $n \in \nn$
the compression map ${\cal C}^{(n)}$ is given by
\begin{eqnarray*}
{\cal C}^{(n)}(\rho^{(n)})=P_{{\cal P}_{\eps_{n}}^{(n)}}\rho^{(n)}P_{{\cal
    P}_{\eps_{n}}^{(n)}} + \sum_{e \in S^{(n)}}| 0 \rangle \langle e | \rho^{(n)}
| e \rangle \langle 0 |, 
\end{eqnarray*}
where ${\cal P}_{\eps_{n}}^{(n)}$ is a high probability subspace of
$\hr^{\otimes n}$ corresponding to the level $\eps_{n}$, $|0 \rangle \in {\cal P}_{\eps_{n}}^{(n)}$  and
$S^{(n)}$ is an orthonormal system in $({\cal
  P}_{\eps_{n}}^{(n)})^{\bot}$. The decompression map ${\cal D}^{(n)}$ is
just the canonical embedding of ${\cal S}(\hr^{(n)})$ into ${\cal
  S}(\hr^{\otimes n})$.
Using the formula (\ref{F_e_sum}) for $F_{e}$ we obtain
\begin{eqnarray*}
F_{e}(\rho^{(n)}, {\cal C}^{(n)})=|\textrm{tr}\rho^{(n)}P_{{\cal
    P}_{\eps_{n}}^{(n)}}|^{2} + \sum_{e \in
  S^{(n)}}|\textrm{tr}\rho^{(n)}| 0 \rangle \langle e ||^{2} \geq |\textrm{tr}\rho^{(n)}P_{{\cal
    P}_{\eps_{n}}^{(n)}}|^{2}.
\end{eqnarray*}
By definition of high probability spaces
$\textrm{tr}\rho^{(n)}P_{{\cal P}_{\eps_{n}}^{(n)}} \geq 1-\eps_{n}$ for all
$n \in \nn$.  Thus \[|\textrm{tr}\rho^{(n)}P_{{\cal
    P}_{\eps_{n}}^{(n)}}|^{2} \geq (1- \eps_{n})^{2} \geq 1-2\eps_{n}.\]
Recall that $\eps_{n}\searrow 0$ and thus assertion \emph{2)} follows. \\  
\emph{Proof of 3)} 
Let us
define for a density operator $\rho$ on $\hr$ and some integer $d \leq \dim \hr$ 
\begin{eqnarray*}
\eta_{d}(\rho):= \max \{ \textrm{tr}\rho P |\ P \textrm{ projector
  on } \hr,
\textrm{tr} P= d \}.
\end{eqnarray*}
As was proven in \cite{barnum}, for any compression scheme $({\cal C}, {\cal D})$ we have
\begin{eqnarray*}
F_{s}(\rho^{(n)}, {\cal C}^{(n)} \circ {\cal D}^{(n)})< 6 \cdot
\eta_{d^{(n)}}(\rho^{(n)}),\qquad \forall n \in \nn,
\end{eqnarray*}
where $d^{(n)}:=\dim \hr^{(n)}$.
Let $\limsup_{n \to
\infty}\frac{1}{n}\log_{2} d^{(n)}=R({\cal
  C})< s(\Psi)$. Then $\lim_{n \to \infty} \eta_{d^{(n)}}(\rho^{(n)})=0$. Otherwise there would exist a sequence $\{P^{(n)}\}_{n \in \nn}$ of
projectors in $\hr^{\otimes n}$, respectively, with asymptotically
not vanishing expectation values 
$\textrm{tr}P^{(n)}\rho^{(n)}=\eta_{d^{(n)}}(\rho^{(n)})$ and $\lim_{n
  \to \infty} \frac{1}{n}\log_{2}\textrm{tr}P^{(n)}=R({\cal C})<s(\Psi)$. This would be a contradiction to
Proposition \ref{beta=s}.
$\qquad \Box$ \\ \\
\emph{Acknowledgement.}
The authors are grateful to Ruedi Seiler, Rainer Siegmund-Schultze and
Tyll Kr\"uger for helpful discussions and valuable comments on this
paper.\\ \\
This work was supported by the DFG via the SFB 288 ``Quantenphysik und
Differentialgeometrie'' at the TU Berlin.


\begin{thebibliography}{99}





\bibitem{barnum} H. Barnum, C. Fuchs, R. Jozsa, B. Schumacher, General
  Fidelity Limit for Quantum Channels, Phys. Rev. A 54, No 6,
  4707-4711 (1996) 

\bibitem{wir}
I. Bjelakovi\'c, T. Kr\"uger, Ra. Siegmund-Schultze, A. Szko\l a, The
Shannon-McMillan Theorem for Ergodic Quantum Lattice Systems,
math.DS/0207121 

\bibitem{datta}  
N. Datta, Yu. Suhov, Data Compression Limit for an Information Source 
of Interacting Qubits, Quant. Inf. Process.  Vol. 1, No. 4, 257-281 (2002)

\bibitem{davies}
E. B. Davies, Quantum Theory of Open Systems, Academic Press, 
London 1976

\bibitem{fannes}
M. Fannes, A Continuity Property of the Entropy Density for Spin
Lattice Systems, Commun. Math. Phys. 31, 291-294 (1973)
\bibitem{hellwig-kraus1}
K.-E. Hellwig, K.Kraus, Pure Operations and Measurements,
Commun. Math. Phys. 11, 214-220 (1969)

\bibitem{hellwig-kraus2}
K.-E. Hellwig, K.Kraus, Operations and Measurements II, 
Commun. Math. Phys. 16, 142-147 (1970)

\bibitem{petz} 
F. Hiai, D. Petz, The Proper Formula for Relative
 Entropy and its Asymptotics in Quantum Probability,
 Commun. Math. Phys. 143, 99-114 (1991)

\bibitem{horodecki}
M. Horodecki, Limits for Compression of Quantum Information by
Ensembles of Mixed States, Phys. Rev. A 57, 3364-3369 (1998)
 
\bibitem{israel}
R.B. Israel, Convexity in the Theory of Lattice Gases,
Princeton, New Jersey 1979

\bibitem{jozsa}
R. Jozsa, B. Schumacher, A New Proof of the Quantum Noisless Coding
theorem, Journal of Modern Optics, Vol.41, No.12, 2343-2349, (1994)

\bibitem{kraus}
K. Kraus, States, Effects and Operations. Fundamental Notions
of Quantum Theory, Lecture Notes in Physics 190, Springer-Verlag,
Berlin 1983 

\bibitem{niel} 
M. A. Nielsen, I. L. Chuang, Quantum Computation and
  Quantum Information, Cambridge University Press, Cambridge 2000

\bibitem{mosonyi}
D. Petz, M. Mosonyi, Stationary Quantum Source Coding,
J. Math. Phys. 42, 4857-4864 (2001)

\bibitem{ruelle} 
D. Ruelle, Statistical Mechanics, W.A. Benjamin, New York 1969

\bibitem{shields}
P.C. Shields, The Ergodic Theory of Discrete Sample Paths, American
Mathematical Society 1996

\bibitem{uhlmann}
A. Uhlmann, Relative Entropy and the Wigner-Yanase-Dyson-Lieb
Concavity in an Interpolation Theory, Commun. Math. Phys. 54, 21-32 (1977)
\end{thebibliography}
\end{document}